# Interfacial superconductivity and zero bias peak in quasi-one-dimensional $Bi_2Te_3/Fe_{1+y}Te$ heterostructure nanostructures

Man Kit Cheng [a, b], Cheuk Yin Ng [a], Sui Lun Ho [a], Omargeldi Atanov [a], Wai Ting Tai [a], Jing Liang [a, b], Rolf Lortz [a, *], Iam Keong Sou [a, b, *]

[a]Department of Physics, The Hong Kong University of Science and Technology, Clear Water Bay, Hong Kong, China.

[b]William Mong Institute of Nano Science and Technology, The Hong Kong University of Science and Technology, Hong Kong, China

[#]Corresponding authors: phiksou@ust.hk & lortz@ust.hk

**Abstract**

$Bi_2Te_3/Fe_{1+y}Te$ heterostructures are known to exhibit interfacial superconductivity between two non-superconducting materials: $Fe_{1+y}Te$ as the parent compound of Fe-based superconducting materials and the topological insulator $Bi_2Te_3$. Here, we present a top-down approach starting from two-dimensional (2D) heterostructures to fabricate one-dimensional (1D) $Bi_2Te_3/Fe_{1+y}Te$ nanowires or narrow nanoribbons. We demonstrate that the $Bi_2Te_3/Fe_{1+y}Te$ heterostructure remains intact in nanostructures of widths on the order of 100 nm and the interfacial superconductivity is preserved, as evidenced by electrical transport and Andreev reflection point contact spectroscopy experiments measured at the end of the nanowire. The differential conductance shows a similar superconducting twin-gap structure as in two-dimensional heterostructures, but with enhanced fluctuation effects due to the lower dimensionality. A zero-bias conductance peak indicates the presence of an Andreev bound state and given the involvement of the topological $Bi_2Te_3$ surface state, we discuss a possible topological nature of superconductivity with strong interplay with an emerging ferromagnetism due to the interstitial excess iron in the $Fe_{1+y}Te$ layer, developing in parallel with superconductivity at low temperatures.

## 1. Introduction

Topological superconductors (SC), especially in the one-dimensional (1D) form, are of great fundamental and technological interest due to their unconventional characteristics. The focus of interest is their potential to accommodate Majorana modes. Localized Majorana bound states promise a topological revolution in quantum computation that is more fault-tolerant [1,2] and exploits their non-Abelian exchange statistics [3,4]. The Majorana bound state or Majorana Zero Mode (MZM) [5] is predicted in various device structures including SC / topological insulator heterostructures [6] and SC/semiconductor heterostructures [7,8]. In quasi-1D nanostructure devices, a $\frac{2e^2}{h}$ quantized conductance [9] in the form of a zero-bias conductance peak (ZBCP) in Andreev reflection experiments at the ends of the nanowires and nonlocal tunnelling of electrons (spin-dependent Andreev reflection [10]) are considered as evidence for the existence of MZM, which, however, are often difficult to distinguish from non-topological Andreev-bound states [11]. Numerous devices for MZM searches have been presented, including but not limited to InSb nanowires in proximity contact with an NbTiN SC [12], hybrid InSb/Nb nanowire junctions [13-

15], Fe atomic chains on Pb [16], HgTe-based topological Josephson junctions [17], proximity-coupled helical hinge states of Bi(111) films decorated with magnetic Fe clusters [18], and superconducting gold surfaces covered by magnetic islands [19,20]. These devices exhibited some of the characteristic physical properties of MZM, although uncertainties about the origins of the claimed signatures were reported for many of them [11,21-23]. Today, the existence of MZM in none of these systems can be considered proven, and further research is needed, including new promising platforms.

Recently, we reported that SC occurs at the interface of a $Bi_2Te_3$/$Fe_{1+y}Te$ heterostructure in which both the $Bi_2Te_3$ and $Fe_{1+y}Te$ layers are non-superconducting. The interface of the $Bi_2Te_3$/$Fe_{1+y}Te$ heterostructure exhibits all features of 2D SC with a Berezinski-Kosterlitz-Thouless (BKT) transition [24]. The observation of a robust 2D vortex system [25] and strong anisotropic magnetic responses as a result of the interplay between superconductivity and the magnetism of interstitial Fe in the $Fe_{1+y}Te$ layer [26], further confirm the 2D SC character. In a series of point-contact experiments, we uncovered a complex superconducting nature with a large pseudogap and a superconducting twin-gap structure associated with the superconductivity at the interface and proximity-induced superconductivity in the $Bi_2Te_3$ layer, respectively [27]. In addition, a zero bias conductance peak (ZBCP) occurred that was robust to high magnetic fields applied perpendicular and parallel to the basal plane. Its origin remains unclear, although the discovery of a SC TI material naturally raises hopes for new topological superconducting states. In this article, we present an approach to study this heterostructure material in the quasi-1D limit, where a magnetic field applied in parallel to a nanowire/nanoribbon structure is expected to turn such proximity-coupled SC TI nanowires into the topological regime [28]. Our motivation is to present a new potential platform for topological superconductivity based on nanostructures fabricated in a relatively simple top-down process using thin-film wafer technology, while revealing a highly complex interplay of magnetic, structural and superconducting degrees of freedom that is of interest alongside the quest for Majorana physics. We will describe the fabrication process and the resulting surface morphology and show that the interfacial superconductivity is preserved, which includes a ZBCP in the differential conductance of Andreev reflection or tunneling experiments at the ends of the 1D nanostructures. We also demonstrate a possible interplay between an emerging ferromagnetism of interstitial Fe in the $Fe_{1+y}Te$ layer and the ZBCP, which could play a role in driving the system into the topological regime.

## 2. Results and Discussion

### 2.1. Device characterization

The nanowire and nanoribbon fabrication is described in detail in the Methods section. Device 1 is a nanowire with Pt electrodes deposited at both ends for Andreev reflection point contact spectroscopy experiments. The width of the $Bi_2Te_3$/$Fe_{1+y}Te$ interface of the nanowire is derived using AFM depth profiling. The corresponding AFM images and data analysis are provided in Fig. 1. The 3D AFM scan in Fig. 1(a) shows that the surface of the nanowire is flatter than the pristine surface with a roughness of a few nanometers. From the height profiles obtained in different regions of Fig. 1(b) (an example is shown in Fig. 1(c) and (d)), the width of the $Bi_2Te_3$/$Fe_{1+y}Te$ interface of the nanowire varies from 80 nm to 120 nm, and the topmost 10 nm ± 3nm were removed during ion milling. The AFM results demonstrate that the $Bi_2Te_3$ layer with a thickness of 6 - 9 QL is well connected along the nanowire, so the $Bi_2Te_3$/$Fe_{1+y}Te$ interface should also exist along the entire length of the nanowire.

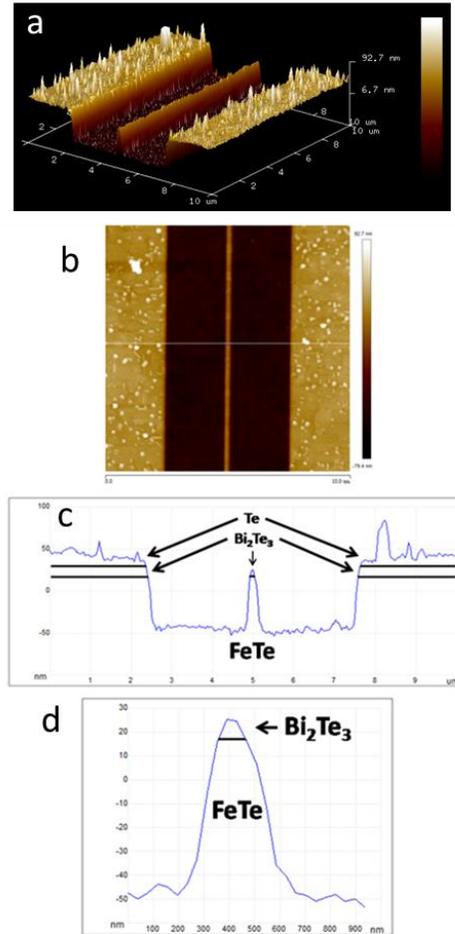

**Figure 1.** 10 μm × 10 μm AFM scans of the $Bi_2Te_3/Fe_{1+y}Te$ nanowire: (a) 3D scan; (b) 2D scan and (c) height profile along the white line marked in (b). (d) Enlargement of the height profile of the $Bi_2Te_3/Fe_{1+y}Te$ nanowire.

For device 2, we wanted to avoid an external contact to the $Bi_2Te_3/Fe_{1+y}Te$ interface with a metal electrode and instead create an internal tunnel contact within a nanoribbon between a bare $Fe_{1+y}Te$ region and a region where the $Fe_{1+y}Te$ was covered by $Bi_2Te_3$. To do this, we simply use FIB to cut a 100 nm wide nanoribbon out of the 2D thin film sample without removing the top layer.

## 2.2. Magnetization and Electrical Resistance

Fig. 2 shows the magnetization of the 2D $Bi_2Te_3/Fe_{1+y}Te$ heterostructure measured on a piece cut from the same thin-film sample used to fabricate devices 1 & 2. The magnetization measured under zero-field cooled (ZFC) and field-cooled (FC) conditions shows a pronounced upturn below 6 K. At the same time, a splitting of the ZFC and FC branches occurs, with the ZFC data lower than the FC data, as indicated by the data showing the difference between the ZFC and FC data, becoming negative below 6 K. An emerging positive magnetization below the superconducting transition temperature is known as the paramagnetic Meissner effect and has been attributed here to the interplay between superconductivity and the strong localized Fe moments of the interstitial Fe in the $Fe_{1+y}Te$ layer [26]. Note that $Fe_{1+y}Te$ does not exist in its stochiometric form and always contains an excess of Fe. We have also added data of the remanent magnetization obtained by first exposing the sample to a high magnetic field of 7 T at the lowest temperature of 1.8 K, then removing the field and measuring the remanent magnetization as the temperature is increased. Below 7 K, a similar upturn to the FC data is seen, suggesting that ferromagnetic correlations emerge in parallel with superconductivity, likely because the strong local field near the interstitial Fe centers generates spontaneous vortices in the superconducting host. The vortices in turn trigger a parallel alignment of the interstitial Fe moments, since the strong anisotropy of the 2D superconductor only allows currents within the 2D plane [25]. The inset in Fig. 2 shows a magnetization loop measured at $T = 1.8$ K, demonstrating a superparamagnetic response known from fine ferromagnetic powder samples [29], here attributed to small clusters of interstitial Fe.

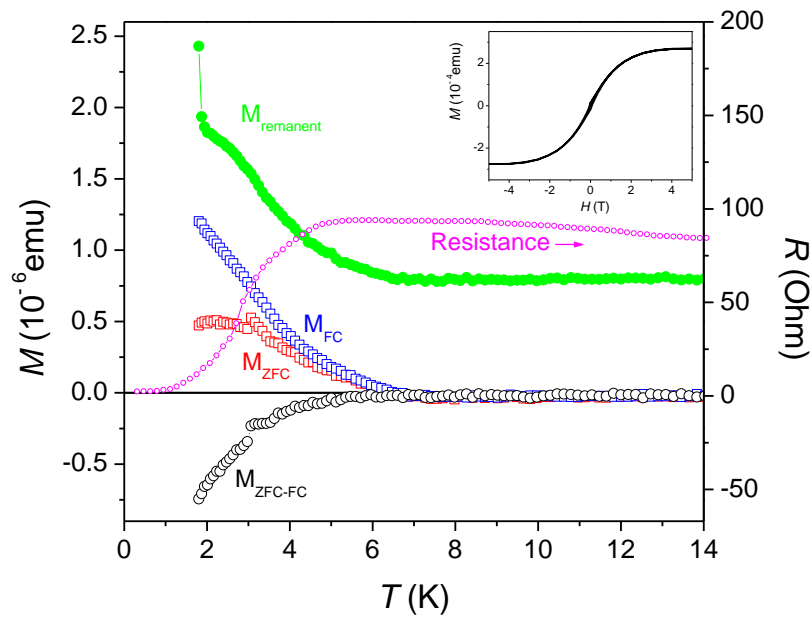

**Figure 2.** Magnetization of a 2D $Bi_2Te_3/Fe_{1+y}Te$ heterostructure measured on a piece cut from the same film used to fabricate device 1 & 2. Shown is the magnetization measured after zero-field (ZFC, red) and field cooling (FC, blue) in a 25 Oe applied field, showing the characteristic positive paramagnetic Meissner effect caused by the interaction of interstitial Fe and superconductivity [26]. Black shows the difference between the ZFC and FC branches, where an ordinary negative Meissner contribution is isolated. Green data shows the remanent magnetization measured during heating in zero field after the sample was magnetized at $T=1.8$ K in a 7 T field (a positive offset of $8 \times 10^{-7}$ emu was added for clarity). Magenta data (right scale): Temperature-dependent zero-field resistance in a 110 nm wide $Bi_2Te_3/Fe_{1+y}Te$ nanowire device (device 1) showing a superconducting transition initiating below 5.5 K. Inset: Magnetization hysteresis loop of the 2D $Bi_2Te_3/Fe_{1+y}Te$ heterostructure measured at $T=1.8$ K. A linear diamagnetic contribution from the substrate was fitted in the high-field saturation region and removed.

We have also added electrical resistance data for device 1. A drop in resistance initiates at 5.5 K, indicating the onset of the superconducting transition, which is consistent with the drop in the data representing the difference between the ZFC and FC data for the 2D heterostructure. The total resistance is a series combination of the resistance of $Bi_2Te_3/Fe_{1+y}Te$ nanowire and the resistance of the regions connecting the nanowire and the bonding pads. Thus, a residual resistance of 235

Ohm originates from the series contact resistance, which is due to the nature of the two-point probe technique used in this study and has been removed in Fig. 2.

## 2.3. Andreev reflection point contact spectroscopy experiments

In Fig. 3a), we show ARPCS data of device 1. The differential conductance at 300 mK as a function of bias voltage of the $Bi_2Te_3/Fe_{1+y}Te$ nanowire device is shown for different magnetic fields applied parallel to the nanowire, in addition to zero-field data at 4.2 K (slightly below the superconducting onset temperature) and 31.5 K (normal state). The 300 mK spectra show a strong positive Andreev reflection signal extending to very high bias voltages above 40 mV, consistent with our previous data on 2D heterostructures [27], where a very large pseudogap was observed in addition to a superconducting twin gap structure. At 4.2 K, the Andreev contribution is still visible, albeit with much smaller amplitude and significantly broadened, while the normal state data at 31.5 K are completely flat, as expected for a metallic material with ohmic characteristics. An applied magnetic field slightly decreases the overall conductivity at 300 mK, but has little effect on the Andreev contribution, since $H_{c2}$ is much higher [24]. This is consistent with ARPCS experiments on 2D heterostructures [27], where for parallel or perpendicular fields up to 15 T, only a small effect of the field was found.

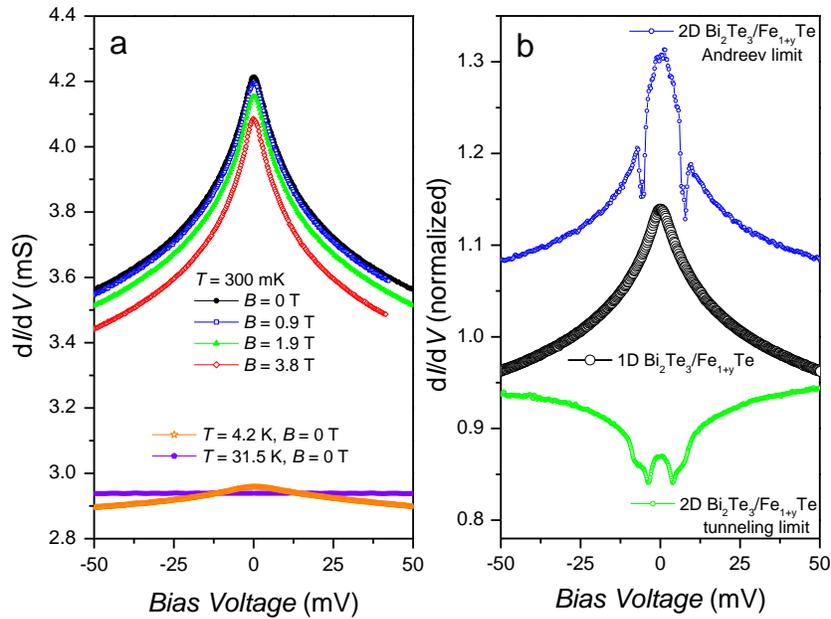

**Figure 3.** Point contact Andreev reflection spectroscopy measured at the end of a 110 nm wide $Bi_2Te_3/Fe_{1+y}Te$ nanowire. a) Differential conductance $dI/dV$ as a function of bias voltage showing a pronounced Andreev reflection signal at 300 mK over a wide bias voltage range, which is hardly influenced by magnetic fields up to 4 T. The Andreev reflection peak is still visible at 4.2 K, while in the normal state at 31.5 K $dI/dV$ is constant. b) Normalized zero-field $dI/dV$ data at 300 mK of the 1D $Bi_2Te_3/Fe_{1+y}Te$ nanowire (central black data). A comparison with point contact spectroscopy data on 2D $Bi_2Te_3/Fe_{1+y}Te$ heterostructures [27] (blue data, high-transparency Andreev limit, green data, low-transparency tunneling limit, scaled up for clarity) shows that the central peak in the 1D nanowire data agrees with the ZBCP observed previously in 2D $Bi_2Te_3/Fe_{1+y}Te$ heterostructures.

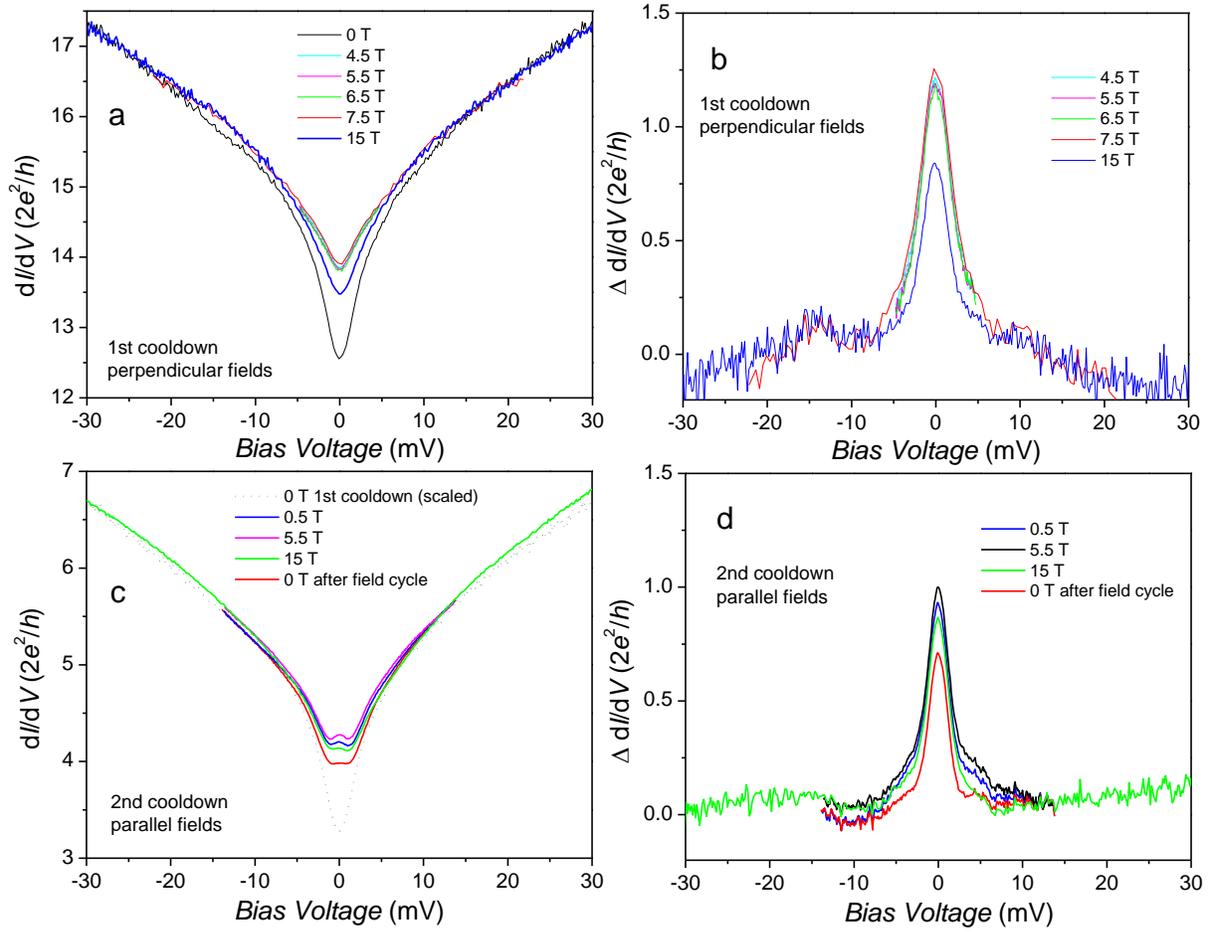

**Figure 4.** Point-contact tunneling spectroscopy measured at the interface between a bare $Fe_{1+y}Te$ and a $Bi_2Te_3/Fe_{1+y}Te$ region within a 100 nm wide nanoribbon (device 2). a) Differential conductance d$I$/d$V$ as a function of bias voltage in zero field and various magnetic fields applied perpendicular to the nanoribbon heterostructure measured after initial cooldown. A V-shape superconducting gap is observed with edges extending to high bias voltages. b) The same d$I$/d$V$ data as in (a), but with the zero-field data subtracted. The difference shows that the zero-field gap is filled in with a ZBCP that reaches a maximum heigh at 7.5 T and decreases towards 15 T at higher fields. c) Differential conductance d$I$/d$V$ as a function of bias voltage after the second cooldown in various magnetic fields applied parallel to the nanoribbon. Before the experiment, the device was exposed to a high magnetic field. A ZBCP is visible inside the gap. The 0T data from the first cooldown, scaled to account for an increase in the tunneling barrier height, was added for comparison. d) The same d$I$/d$V$ data as in (c), but with the scaled zero field data from the first cooldown subtracted, showing the evolution of the zero-bias peak reaching a maximum at 5.5 T and remaining partially in zero field after the field cycle.

Fig. 4a-d shows the point contact spectroscopy data for device 2. When the device was first cooled, the $Bi_2Te_3/Fe_{1+y}Te$ film plane and the nanoribbon direction were oriented perpendicular to the magnetic field direction. Fig. 3a shows the differential conductance d$I$/d$V$ as a function of bias voltage at 0T and different magnetic fields after the first cooling. In contrast to device 1, the tunneling interface had a high barrier height, so that the data were dominated by tunneling processes. A pointed superconducting gap is observed with edges extending to similar high bias

voltages as in device 1. In a magnetic field, the gap initially fills up to 7.5 T, but deepens again in the highest field of 15 T. This cannot be explained by the usual effect of the magnetic field closing the superconducting gap. It suggests that the gap is rather filled by an Andreev-bound state, which reaches a maximum at intermediate fields and weakens at 15 T, although the ZBCP is obscured by the acute depression of the superconducting gap. Note that hardly any field effect was observed in tunneling experiments on 2D $Bi_2Te_3$/$Fe_{1+y}Te$ heterostructures due to an extraordinarily high upper critical field up to 15 T [24]. Therefore, in Fig. 4b we show the same d$I$/d$V$ data in magnetic fields with the zero-field data subtracted as background. The difference confirms that the zero-field gap is filled with a ZBCP that reaches a maximum height at 7.5 T and decreases towards 15 T at higher fields.

In Fig. 4c we plot the differential conductance d$I$/d$V$ as a function of bias voltage after the second cooldown in different magnetic fields, this time applied in-plane and parallel to the nanoribbon. Note that the contact resistance increased slightly during the thermal cycle to room temperature and back, so that the overall conductance decreased. Unfortunately, prior to the experiment, the device was subjected to a high magnetic field as another experiment was performed in the same probe, so the initial state in the zero field immediately after cooling is unknown. The overall shape of the data at higher bias is very similar to the data measured after the first cooling, which is not surprising since the device was already well into the high barrier tunneling limit, even though the tunnel barrier continued to increase after the second cooling. We scaled the zero-field data from the first cool-down to the same value at high bias and added it for comparison, which matches the new data perfectly. However, a small ZBCP is now visible within the gap, increasing up to 5.5 T, while the gap acquires a flat central region in zero field following the field cycle, unlike the first cool-down data. Fig. 4d shows the same d$I$/d$V$ data in magnetic fields, but with the scaled zero field data from the first cool-down cycle subtracted as background, showing the evolution of the ZBCP reaching a maximum at 5.5 T and remaining partially in the zero field after the field cycle.

## 2.4. Discussion

We have previously reported that 2D $Bi_2Te_3$/$Fe_{1+y}Te$ heterostructures with interfacial superconductivity exhibit a very large superconducting twin-gap structure, which was replaced by a pseudogap above ~12 K that persisted up to 40 K [27]. In the low-transparency Andreev limit, they merged into a broad positive peak with some shoulders caused by the gap edges. The larger gap was attributed to a thin $Fe_{1+y}Te$ layer in proximity to the interface, and the smaller gap was associated with superconductivity induced by the proximity effect in the $Bi_2Te_3$ topological insulator. The pseudogap was interpreted because of the strong fluctuations in a 2D superconductor in which the resistive transition occurs in the form of a Berezinskii-Kosterlitz-Thouless phase-ordering transition [24] at much lower temperature than Cooper pair formation. No shoulders or dips can be resolved in our data of device 1 and 2 for the 1D nanowires, which, as we will show below, is due to reduced quasiparticle lifetime due to enhanced fluctuations in this even lower dimensionality. However, the nanostructured devices show a central zero-bias peak, which is consistent with the ZBCP previously observed in 2D $Bi_2Te_3$/$Fe_{1+y}Te$ heterostructures, but for our 1D nanowires/nanoribbons the ZBCP is sharper.

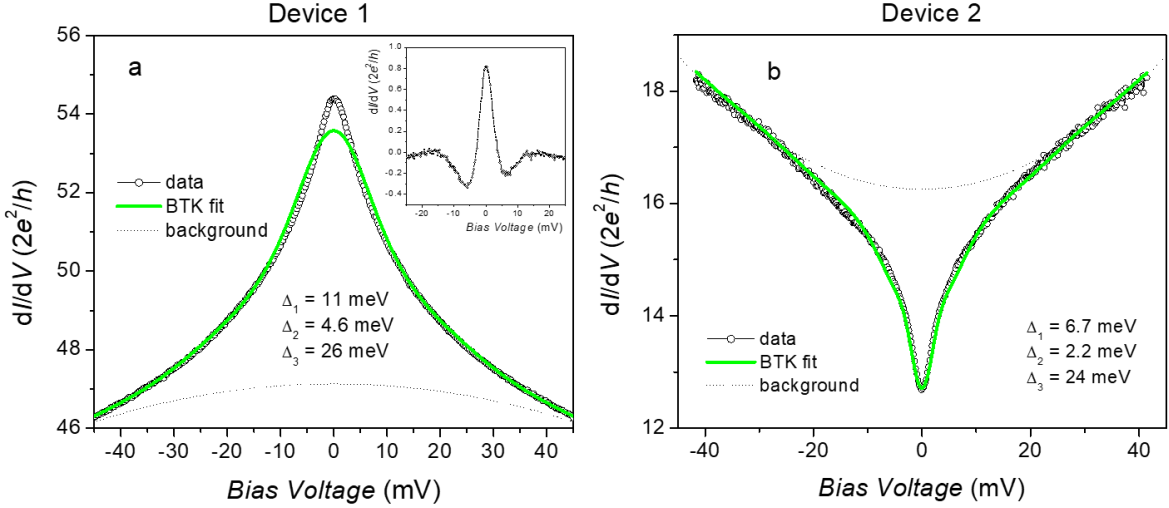

**Figure 5.** Zero-field d$I$/d$V$ data measured at 300 mK after the first cooldown of device 1 (a) and 2 (b) with a modified Blonder-Tinkham-Klapwijk (BTK) fit (green line) using a three-band model with $\Delta_1$=11 meV, $\Delta_2$=4.6 meV and $\Delta_3$=26 meV (device 1) or $\Delta_1$=6.7 meV, $\Delta_2$=2.2 meV and $\Delta_3$=24 meV (device 2). A smooth parabolic background is used to account for the curvature at high bias (dotted line). For device 1, the zero bias region was excluded to account for a ZBCP. The inset in (a) shows the difference between device 1 data and the fit that shows the ZBCP more clearly.

Note that the in-plane superconducting coherence length is shorter than the width of our nanostructures [24], but it has been estimated that for $Bi_2Te_3$ related TI materials, 100 nm is a critical width for hybridization of the edge states [30], which means that the TI can be considered to be within the 1D limit.

We analyse the zero-field $Bi_2Te_3$/$Fe_{1+y}Te$ nanowire data by fitting the differential conductance versus bias voltage with a modified 1D Blonder-Tinkham-Klapwijk (BTK) model considering a finite-transparency tunnel junction [31]. Since we are dealing with a low-dimensional system, eventually in the crossover region between the 2D and 1D case, fluctuation effects are expected to be enhanced, leading to short quasiparticle lifetimes, which we account for by a large Dynes parameter [32]. To account for the sharp ZBCP in device 1, we exclude a region on the order of 2-3 meV around zero bias, while for device 2 we use the zero-field data measured following the initial cooldown, which does not have a ZBCP. In addition, we used an inverted parabolic background to account for the curvature in the bias voltage range above 30 mV (dotted lines in Fig. 4. The third gap is added to account for the large pseudogap previously observed in 2D heterostructures [27].

The results of the fits are shown in Fig. 4. The model describes the data very well. For device 1, the two main superconducting gaps of $\Delta_1$=11 meV and $\Delta_2$=4.6 meV are only slightly smaller than the values in 2D heterostructures (12 meV and 6 meV, respectively) [27], which is likely due to the slightly lower $T_c$ value. With the large Dynes parameters, the structures caused by the 3 gaps merge together to account for the very broad range of the observed Andreev peak of characteristic λ shape in device 1, or the gap structures in device 2, without any shoulders due to the edges of the gaps. The ZBCP in device 1 cannot be accounted for unless an unrealistically small fourth gap were added to the model. The sharpness of the zero-bias anomaly in device 1 shows that a ZBCP is also present here, even if it merges with the main Andreev peak formed by $\Delta_1$ and $\Delta_2$. To account for the presence of the ZBCP, we also tried to reproduce these data with a two-gap *d*-wave model, but no improvement in the fit or significant change in gap values was observed compared to the *s*-wave case, although a nodal superconducting order parameter would naturally explain the presence of a ZBCP [33,34]. For device 2 the gap values are smaller: $\Delta_1$=6.7 meV and $\Delta_2$=2.2 meV. The reason for this difference is unclear, but ultimately the device is even more one-dimensional, which weakens the interfacial superconductivity.

Our temperature-dependent data for device 1 indicate that the gaps are quickly smeared and form only a broad bump at 4 K, while well above $T_c$ at 31.5 K the differential conductance is completely constant, indicating the metallic normal state with ohmic current-voltage characteristics.

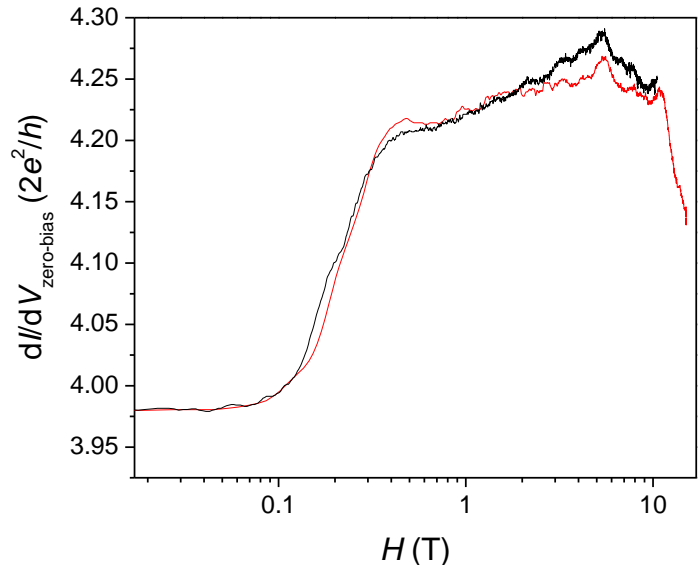

**Figure 6.** Differential conductance d*I*/d*V* of device 2 measured at zero bias as a function of the magnetic field applied parallel to the $Bi_2Te_3$/$Fe_{1+y}Te$ nanoribbon in two repeated measurements to demonstrate reproducibility. Between 0.1 T and 0.3 T, there is a sharp increase in the conductance, which is consistent with the sudden increase in the ZBCP in Fig. 4 c and d.

In our previous report [27], we discussed some possible causes of the ZBCP in 2D heterostructures. It should be noted here that currently neither the superconducting pairing symmetry nor the pairing mechanism of interfacial superconductivity in $Bi_2Te_3$/$Fe_{1+y}Te$ heterostructures is known. Since the parent compound $Fe_{1+y}Te$ of Fe-based high-temperature superconductors is one of the building blocks of this heterostructure, an unconventional nature of superconductivity is likely. Angular photoemission spectroscopy revealed that the superconductivity is possibly induced by charge transfer across the interface, suggesting a hole-doped FeTe region that likely triggers unconventional superconductivity [35]. In addition, there is the involvement of the topological surface state of $Bi_2Te_3$ at the interface [24,36,37], superconductivity propagating into the topological insulator layer due to the proximity effect [27], and the presence of local magnetism from the interstitial excess iron in the $Fe_{1+y}Te$ layer [26]. A nodal order parameter such as p- or d-wave could naturally account for a ZBCP. The case of unconventional superconductivity on a topological insulator was discussed in Ref. 38. For a topological system, it is expected that for the s-wave case a finite magnetic field is required to induce zero-energy states, while in device 1 the ZBCP is already present in zero field. On the other hand, in the case of a $d_{xy}$-wave pairing symmetry, for example, a ZBCP is naturally formed in zero field due to resonant states in the middle of the gap. Within a superconducting topological insulator, zero energy states are not spin degenerate, causing a Majorana nature of the fermions in contrast to a topological trivial $d_{xy}$-wave superconductor where the zero energy modes are spin degenerate. When a Zeeman field is introduced, the ZBCP may split or become suppressed for certain directions. Our results, including those from Ref. 27, do not indicate any strong field dependence and rule out splitting of the ZBCP. Scanning tunnelling experiments on FeSe and $FeSe_{1-x}Te_x$ have found evidence of nodal symmetry only for stoichiometric FeSe by vortex core spectroscopy [39]. After substitution with Te, it was suspected to become a nodeless $s^{\pm}$ superconductor. This is probably also true for $Bi_2Te_3/Fe_{1+y}Te$, as our data for device 2 showed no ZBCP immediately after initial cooling, ruling out nodal order parameter symmetry as a cause. A magnetic field was required to generate it in this device. Moreover, the low height near the quantization suggests that the ZBCP is related to the smallest gap $\Delta_2$, which is due to the superconductivity induced by the proximity in the $Bi_2Te_3$ layer [27].

To investigate the effect of a magnetic field on the ZBCP in more detail, we measured the differential conductance for device 2 after the second cooldown at zero bias voltage as a function of the magnetic field applied parallel to the nanoribbon. A sudden, reproducible increase in the height of the ZBCP is observed between 0.1 T and 0.3 T. Beyond 0.3 T, the height continues to increase and peaks at 5.5 T, where we also observed the largest height in Fig. 4d. This field dependence suggests another possible explanation for the ZBCP, which could be related to the excess interstitial iron in $Fe_{1+y}Te$ [33] causing Andreev bound states in the gap. It has been shown that odd-frequency pairing amplitude arises in spatially non-uniform superconductors, attributed here to the excess iron, which can lead to zero-energy Andreev bound states at the surface or interfaces of the superconductor [40]. The fact that superconductivity occurs at the interface with a topological insulator, naturally raises the hope that Majorana physics may be involved in such a zero-energy Andreev bound state. On the surface of a topological insulator like $Bi_2Te_3$ there is only one surface Dirac fermion, and in presence of superconducting pairing Majorana modes will appear. When a ferromagnetic insulator is brought into contact with the proximity coupled topological insulator, the emergence of a one-dimensional chiral Majorana mode was predicted as an Andreev bound state, in addition to the MZMs that form in the vortex cores [40-43]. Therefore, there are possible scenarios with MZMs that could apply to our case. For 2D $Bi_2Te_3/Fe_{1+y}Te$ heterostructures, one would expect localized Majorana zero modes within the vortex cores [41,44]. However, the excess Fe forms strong local magnetic moments with some ferromagnetic exchange at low temperatures, which should lead to spontaneous vortex formation without the need for externally applied magnetic fields [26]. This is supported by the paramagnetic Meissner effect we observed in our magnetization data (Fig. 2), which appears below the onset of superconductivity and suggests a strong interplay between superconductivity and the local Fe moments. If such a spontaneous vortex is formed near the point contact, it can cause a ZBCP. Otherwise, clusters of interstitial Fe may directly cause the one-dimensional chiral Majorana mode described above. In a 1D nanowire geometry in proximity to an s-wave superconductor, a field applied parallel to the nanowire axis is expected to drive the material into the topological regime [28], with a pair of localized MZM forming at the opposite ends. Here, the interstitial Fe could provide a local field that allows us to see a ZBCP even without externally applied field. This is supported by the dependence of the height of the ZBCP in fields applied parallel to the nanoribbon, which align the magnetic Fe moments along the nanoribbon direction. For a detailed understanding of the ZBCP, it is therefore probably necessary to investigate the role of the interstitial Fe clusters in the $Fe_{1+y}Te$ layer on the superconductivity in more detail.

## 3. Conclusions

We have presented a fabrication process for $Bi_2Te_3/Fe_{1+y}Te$ nanowire devices for electrical transport and Andreev reflection point contact spectroscopy experiments at one end of the nanowire. Atomic force microscopy was used to determine the surface morphology of the $Bi_2Te_3/Fe_{1+y}Te$ nanowire, demonstrating that the heterostructure remains intact in the nanowire. Electrical transport and Andreev reflection point contact spectroscopy show that the interfacial superconductivity is preserved and exhibits similar characteristics to 2D $Bi_2Te_3/Fe_{1+y}Te$ heterostructures with a twin-gap structure and a zero-bias conductance anomaly. While the overall gap structures appear rather smeared due to enhanced fluctuations in the reduced dimensionality, the ZBCP is sharper than in 2D samples and we have discussed possible links to unconventional superconductivity, Andreev bound states from the excess iron in $Fe_{1+y}Te$ or Majorana zero modes. With the present experiment, we are only at the beginning and cannot provide a conclusive explanation for the ZBCP. The search for Majorana zero modes has been proven to be far more difficult than originally assumed, with the most spectacular reports on its observation all been retracted. The aim of this work is to promote a new potential platform, which can be easily realized in a simple top-down approach, and as we demonstrated with our device 2, with a particularly clean tunneling interface formed within the epitaxially-grown material without the need for an external tunneling lead, thus avoiding the so-called "soft-gap" problem [45], which is considered the main obstacle in approaches based on free-standing nanowires [46]. Further experiments should ideally be performed on heterostructures with higher $T_c$ [24] and at lower temperatures in a dilution refrigerator so that lower reduced temperatures $T/T_c$ are accessible to reduce fluctuations and obtain sharper features. An interesting tool would be microscopic experiments, e.g. with a scanning tunneling probe, to see if there is a connection between the ZBCP and the excess iron. Furthermore, it would be interesting to deposit some nanoscale Fe clusters near the interface to study the effect of the exchange field in a more controlled way [16,17].

## 4. Experimental Section

*Device fabrication:* The $Bi_2Te_3/Fe_{1+y}Te$ nanowire device was fabricated from a $Bi_2Te_3/Fe_{1+y}Te$ thin film heterostructure sample grown in a VG V80H MBE system. The structure of the $Bi_2Te_3/Fe_{1+y}Te$ thin film sample is Te (10nm)/$Bi_2Te_3$ (8-9nm)/$Fe_{1+y}Te$ (220nm)/ZnSe (80nm)/$n^+$ GaAs (001). The ZnSe layer serves as buffer layer and Te as a capping layer for protection. The thin film sample was cut into a small piece with an area of 2mm × 4mm and transferred into an FEI Helios G4 UX dual-beam Focused Ion Beam (FIB) / Field Emission Scanning Electron Microscope (FESEM) system for ion milling and Pt electrode deposition was applied to fabricate the $Bi_2Te_3/Fe_{1+y}Te$ nanowire and the corresponding device for electrical transport measurements. Scanning Electron Microscope (SEM) images were acquired using the dual-beam FIB/FESEM system. Surface morphology with depth profile of the $Bi_2Te_3/Fe_{1+y}Te$ nanowire was obtained using a Bruker Dimension ICON Atomic Force Microscope (AFM) with a ScanAsyst-Air probe, and the results were analyzed by NanoScope Analysis software. Ti(5nm)/Au(70nm) bonding pads were deposited onto the device using a Denton Vacuum Desk Top Pro desktop sputtering system. Electrical contacts were made with Al wires bonded by an ASM AB520 wedge bonder. High Resolution Transmission Electron Microscopy (HRTEM) images were captured using a JEOL

JEM 2010F TEM.

Fig. 7 shows a schematic diagram of the fabrication process of $Bi_2Te_3/Fe_{1+y}Te$ nanowires for electrical measurements with two probes (device 1). The thicknesses of the different layers of the thin film sample were measured using the HRTEM images shown in Fig. 8. The $Bi_2Te_3/Fe_{1+y}Te$ nanowires were fabricated in the dual-beam FIB/FESEM system using a $Ga^+$ ion beam source at 30 keV. The ion beam source current ranged from 1 pA to 65 nA for different aspects. A high surface flatness region was used to fabricate the $Bi_2Te_3/Fe_{1+y}Te$ nanowire, which was investigated using tilted SEM and FIB images. The FIB source current for image acquisition was set to 1 pA to reduce damage to the sample. Optimized currents were selected for ion milling at different scales. Then, a 30 μm long $Bi_2Te_3/Fe_{1+y}Te$ nanowire was fabricated by ion milling and the nearby $Bi_2Te_3$ layer was removed to ensure that the SC properties were dominated by the $Bi_2Te_3/Fe_{1+y}Te$ nanowire in transport and differential conductivity measurements. The unetched part near the $Bi_2Te_3/Fe_{1+y}Te$ nanowire is used for height calibration.

The $Bi_2Te_3$ layer near the expected position of the Pt and Ti/Au electrodes was also removed for the reasons mentioned above. The FeTe and ZnSe layers with a width of 5 μm on the two sides near the center of the nanowire were removed up to the two top/bottom edges of the device (Fig. 7) to prevent a short circuit. After the ion milling, Pt wires with a cross-sectional area of 300 × 300 nm were deposited to connect the two ends of the nanowire to the positions of the Ti/Au bonding pads. Then the device was transferred to the sputtering system and the two ends of the device were coated with Ti/Au electrodes. Finally, the device was mounted on a ceramic leadless chip carrier (CLCC) with paraffin wax, and the two Ti/Au bonding pads were connected with the pads on the CLCC by Al bonding wires. Fig. 9(a) and (b) show SEM images of the $Bi_2Te_3/Fe_{1+y}Te$ nanowire.

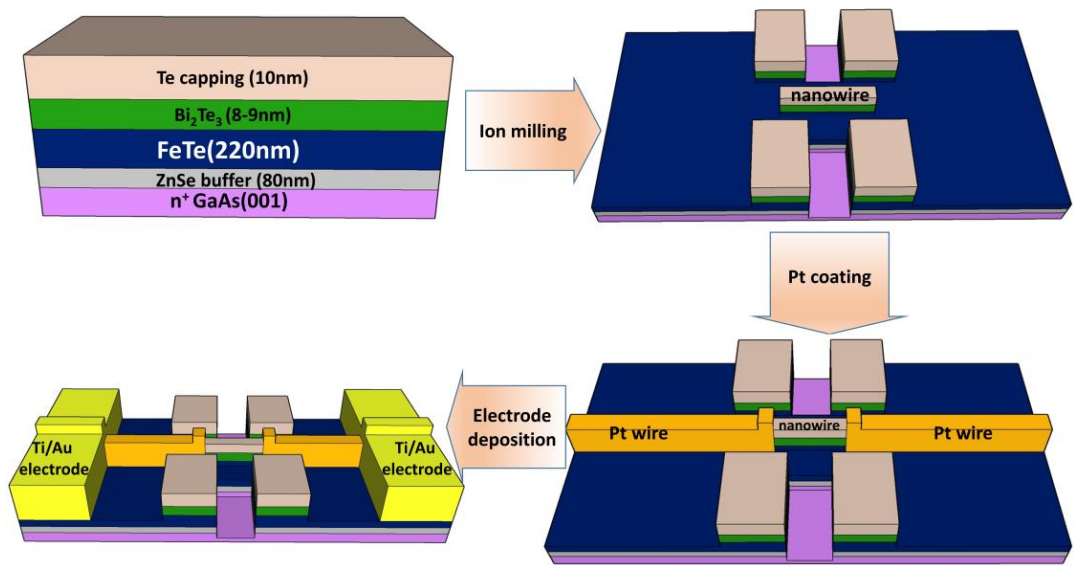

**Figure 7.** Schematic drawings of the fabrication processes of the $Bi_2Te_3/Fe_{1+y}Te$ nanowire device for two-point probe electrical transport and Andreev reflection point contact measurements.

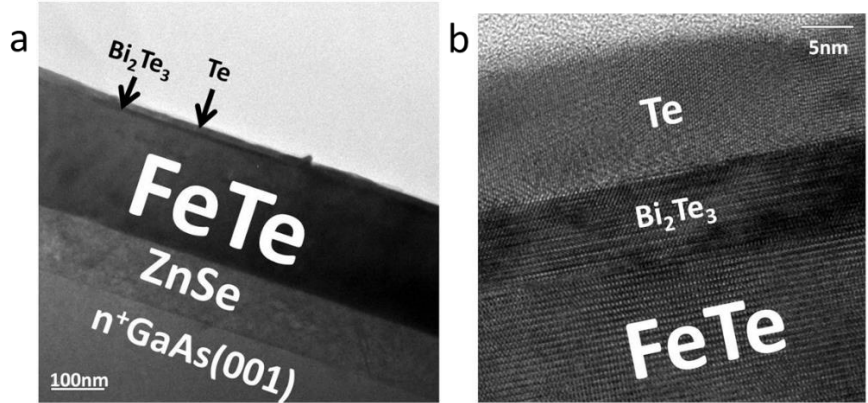

**Figure 8.** The cross-sectional HRTEM image of the $Bi_2Te_3/Fe_{1+y}Te$ thin film sample. (a) The full picture of the sample; (b) close-up image near the $Bi_2Te_3/Fe_{1+y}Te$ interface.

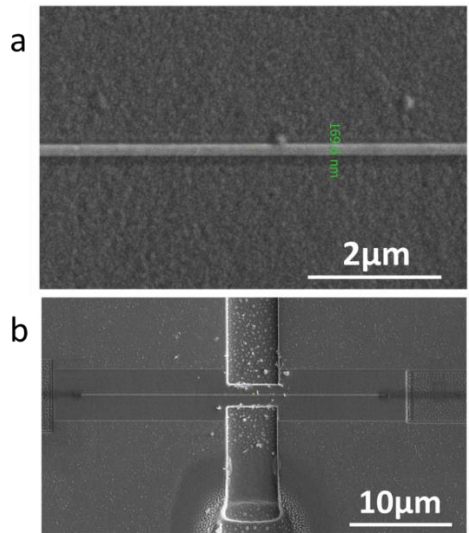

**Figure 9.** The SEM images of (a) the close-up and (b) full picture of $Bi_2Te_3/Fe_{1+y}Te$ nanowire.

For device 2, we wanted to avoid contact with the $Bi_2Te_3/Fe_{1+y}Te$ interface through a metal electrode and instead create an internal tunneling contact within the heterostructure material. The $Bi_2Te_3/Fe_{1+y}Te$ thin film sample used to fabricate the nanoribbon device was the same as that used for device 1. The detailed structure of the nanoribbon device is shown in Fig. 10. The Te capping layer and $Bi_2Te_3$ layer were removed from the central region of the $Bi_2Te_3/Fe_{1+y}Te$ thin film sample by ion sputtering conducted in a IONTOF TOF.SIMS 5 secondary ion mass spectroscopy (SIMS) system. The sample was then transferred to an FIB device for nanoribbon fabrication. Here the FIB was simply used to cut through the Te, $Bi_2Te_3$ and FeTe layers, leaving behind a ~100 nm wide nanoribbon. Half of the nanoribbon consists only of the lowest FeTe layer, which serves as a normal conducting metallic lead, while the other half consists of the entire heterostructure, creating a particular clean, epitaxially-grown tunneling barrier.

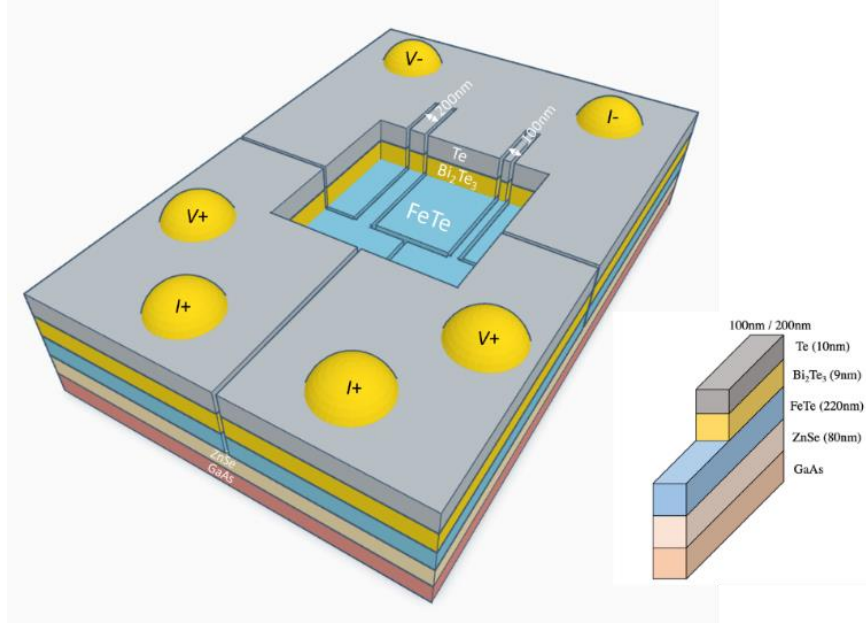

**Figure 10.** a) 3D sketch of the device for point contact spectroscopy at the interface from a normal conducting $Fe_{1+y}Te$ nanoribbon (used as metallic lead) to a $Bi_2Te_3/Fe_{1+y}Te$ heterostructure nanoribbon with interfacial superconductivity fabricated in this study. In the center the top Te and $Bi_2Te_3$ layers have been removed to expose the $Fe_{1+y}Te$ layer. Using focused ion beam techniques, two nanoribbons of 200 nm and 100 nm width have been fabricated crossing the bare $Fe_{1+y}Te$ region in the center to the Te-capped $Bi_2Te_3/Fe_{1+y}Te$ region. Subsequently, 3 electrodes have been separated in the outer region, each with a voltage and current terminal to allow electrical transport measurements across each of the two nanoribbons. b) Cut through the film identifying the different layers including the GaAs substrate, a ZnSe buffer layer, the $Fe_{1+y}Te/Bi_2Te_3$ heterostructure and a Te capping layer.

*DC Magnetization, electrical transport and point contact Andreev reflection experiments:* The DC magnetization of a 2D $Bi_2Te_3/Fe_{1+y}Te$ heterostructure was measured using a commercial Quantum Design Vibrating-sample SQUID magnetometer. Electrical transport and Andreev reflection point contact spectroscopy (ARPCS) measurements were conducted across the quasi-1D nanostructures using a Keithley 6221 AC/DC current source, a 34411A digital multimeter, and an SR830 lock-in amplifier. The electrical transport experiments were performed in AC mode at a frequency on the order of 10 Hz in combination with the lock-in-amplifier. For the point contact spectroscopy, the current source generated an AC current with variable DC offset. The DC component induces the bias voltage by injecting the current in small steps from a negative value to a positive value across the point contact formed by the Pt electrode at one end of the nanowire, while the other side was grounded. The DC bias voltage was measured using the digital multimeter. The AC current with a constant amplitude comparable to the step size of the bias current scan was used to determine the differential conductance $dI/dV$ as a function of the measured bias voltage.

## Acknowledgements

M.K.C., C.Y.N and S.L.H. contributed equally to this work. We thank U. Lampe for technical assistance with the low-temperature experiments. The authors acknowledge funding by the

Research Grants Council of the Hong Kong Special Administrative Region, China, under Grant Numbers 16308020, 16302018, 16302319, C6025-19G, SBI17SC14 and IEG16SC03.

**Conflict of Interest**

The authors declare no conflict of interest.

**Data Availability Statement**

The data that support the findings of this study are available from the corresponding author upon reasonable request.

**Keywords**
topological superconductivity, interfacial superconductivity, quasi-one-dimensional superconductivity, Andreev reflection point contact spectroscopy, Majorana zero mode